\documentclass[journal]{vgtc}                     

\onlineid{0}

\vgtccategory{Research}

\vgtcpapertype{please specify}

\title{LocoScooter: Designing a Stationary Scooter-Based Locomotion System for Navigation in Virtual Reality}

\author{%
  \authororcid{Wei He}{0009-0001-6745-2752},
  \authororcid{Xiang Li}{0000-0001-5529-071X}, \authororcid{Per Ola Kristensson}{0000-0002-7139-871X} and 
  \authororcid{Ge Lin Kan}{0000-0002-4422-9531}
}

\authorfooter{

  \item Wei He and Xiang Li contributed equally to this work.
  \item Wei He and Ge Lin Kan are with The Hong Kong University of Science and Technology (Guangzhou). 
  \\E-mail: whe694@connect.hkust-gz.edu.cn, gelin@hkust-gz.edu.cn.

  \item Xiang Li and Per Ola Kristensson are with the University of Cambridge. 
  \\E-mail: \{xl529\,$|$\,pok21\}@cam.ac.uk\,.

}

\abstract{
\rev{Virtual locomotion remains a challenge in VR, especially in space-limited environments where room-scale walking is impractical. We present LocoScooter, a low-cost, deployable locomotion interface combining foot-sliding on a compact treadmill with handlebar steering inspired by scooter riding. Built from commodity hardware, it supports embodied navigation through familiar, physically engaging movement. In a within-subject study ($N = 14$), LocoScooter significantly improved immersion, enjoyment, and bodily involvement over joystick navigation, while maintaining comparable efficiency and usability. Despite higher physical demand, users did not report increased fatigue, suggesting familiar movements can enrich VR navigation.}
} 


\teaser{
  \centering
  \includegraphics[width=\textwidth]{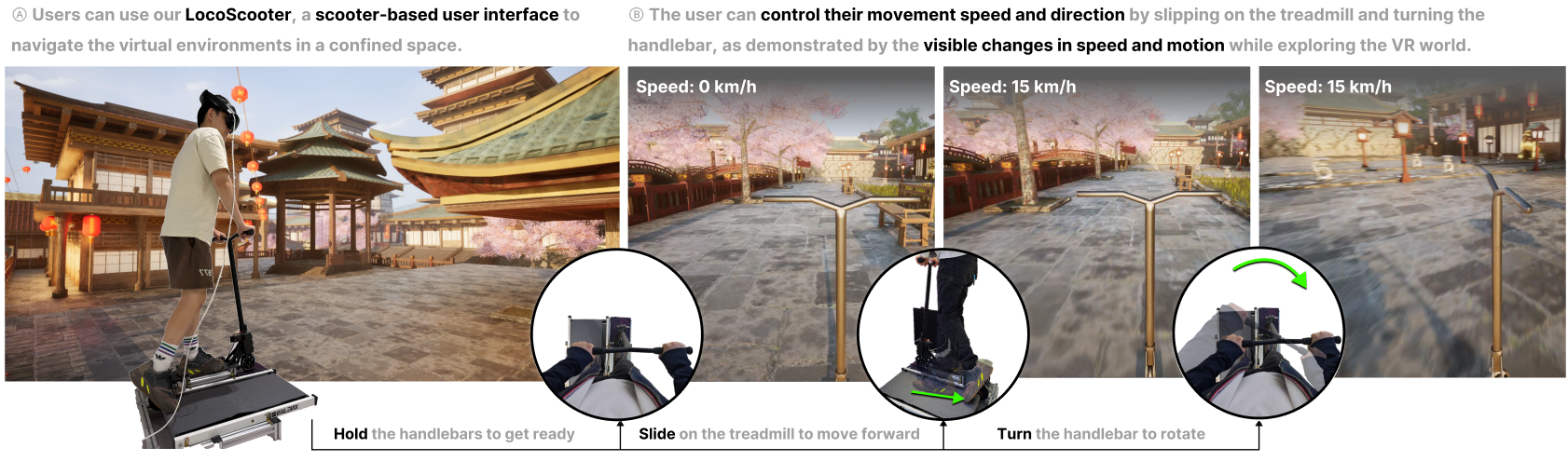}
   \caption{In this paper, we present LocoScooter, a scooter-based embodied locomotion interface for confined virtual environments. Users propel themselves by sliding one foot on the treadmill-like base and steer by rotating the handlebars.}
  \label{fig:teaser}
}

\graphicspath{{figs/}{figures/}{pictures/}{images/}{./}} 

\usepackage{tabu}                      
\usepackage{booktabs}                  
\usepackage{lipsum}                    
\usepackage{mwe}                       

\usepackage{mathptmx}                  

\usepackage{times}                     

\usepackage{tabu}                      
\usepackage{booktabs}                  
\usepackage{lipsum}                    
\usepackage{mwe}      
\usepackage[table]{xcolor} 

\usepackage{xcolor}
\usepackage[normalem]{ulem}

\newif\ifshowrevs
\showrevsfalse 

\ifshowrevs

  \newcommand{\rev}[1]{\textcolor{blue}{#1}}
   \newcommand{\del}[1]{\textcolor{red}{\sout{#1}}}

\else
  \newcommand{\rev}[1]{#1}
  \newcommand{\del}[1]{}
\fi

\definecolor{bad}{RGB}{255,110,110}
\definecolor{mediumbad}{RGB}{255,190,190}
\definecolor{mediumgood}{RGB}{226,239,217}
\definecolor{good}{RGB}{168,208,141}

\begin{document}


\firstsection{Introduction}

\maketitle

Navigating large-scale virtual environments (VEs) remains a fundamental challenge in virtual reality (VR), especially in confined physical spaces \cite{Vasylevska_2017_compressing, Williams_2007_exploring,caporusso2020immersive}. While expansive VEs offer rich and immersive experiences, they often demand locomotion techniques that reconcile limited physical space with naturalistic control. Common approaches such as joystick-based navigation and teleportation provide compact and accessible solutions but lack embodied feedback, leading to reduced presence and spatial coherence due to the disconnect between physical effort and virtual movement \cite{chen_2013_navigation,boletsis2019vr}.

Prior research shows that aligning real-world and virtual body movement, such as through physical walking, can significantly improve presence compared to symbolic or indirect techniques like walking-in-place or flying \cite{usoh1999walking}. However, implementing such physically grounded locomotion in everyday settings is often infeasible due to spatial or hardware constraints. While alternatives like redirected walking, walk-in-place methods, and omnidirectional treadmills \cite{razzaque2005redirected,bruno2013new,wilson2016vr,calandra2018arm,warren2017user} offer more bodily engagement, they typically require complex setups or dedicated spaces, making them unsuitable for casual or shared VR use. This motivates the need for a compact, deployable, and embodied locomotion interface, one that treats the human body as an expressive input source \cite{li2024menu,mueller2023toward}, yet remains lightweight and context-appropriate.

In this paper, we present \textit{LocoScooter}, a scooter-inspired locomotion interface designed for confined VR settings such as educational demos, fitness games, and virtual exhibitions. Users navigate virtual environments by sliding one foot on a treadmill-like platform while steering with handlebars (see Figure~\ref{fig:teaser}), mimicking the embodied dynamics of scooter riding\rev{---specifically, the coordination of asymmetric balance, rhythmic foot propulsion, and upper-body steering}. The system allows continuous, full-body control of both speed and direction within a 0.5\,m\textsuperscript{2} footprint. Built entirely from commodity components, \textit{LocoScooter} offers a low-cost, physically engaging alternative to joystick-based control, without requiring room-scale tracking or dedicated infrastructure.

We evaluated \textit{LocoScooter} in a within-subject user study ($N = 14$), comparing it to joystick-based navigation in a large-scale virtual city task. The task design emphasized directional variety and spatial traversal, calibrated through pilot testing to ensure ecological realism. We assessed task performance, usability, perceived workload and fatigue, simulator sickness, and user experience, supported by semi-structured interviews.

\begin{table*}[t]
\centering
\caption{Comparison across common VR locomotion techniques and LocoScooter along key dimensions (red = poor, green = best).}
\label{tab:locomotion_comparison}
\resizebox{\linewidth}{!}{%
\begin{tabular}{lcccccc}
\toprule
\textbf{Dimension} & \textbf{Teleportation} & \textbf{Smooth Locomotion} & \textbf{Arm Swinging} & \textbf{Room-Scale Walking} & \textbf{Omni-directional Treadmill} & \textbf{LocoScooter} \\
\midrule
\textit{Physical Engagement}       & \cellcolor{bad}No & \cellcolor{bad}No & \cellcolor{good}Yes & \cellcolor{good}Yes & \cellcolor{good}Yes & \cellcolor{good}Yes \\
\textit{Continuous Movement}       & \cellcolor{bad}No & \cellcolor{good}Yes & \cellcolor{good}Yes & \cellcolor{good}Yes & \cellcolor{good}Yes & \cellcolor{good}Yes \\
\textit{Directional Control}       & \cellcolor{mediumgood}Yes & \cellcolor{mediumgood}Yes & \cellcolor{mediumbad}Limited & \cellcolor{good}Natural & \cellcolor{good}Natural & \cellcolor{mediumgood}Yes (handlebar steering) \\
\textit{Speed Modulation}        & \cellcolor{mediumbad}Limited & \cellcolor{mediumgood}Yes & \cellcolor{mediumbad}Limited & \cellcolor{good}Natural & \cellcolor{good}Natural & \cellcolor{mediumgood}Yes (foot sliding) \\
\textit{Motion Congruence}         & \cellcolor{bad}No & \cellcolor{mediumbad}Low & \cellcolor{mediumgood}Partial & \cellcolor{good}Yes & \cellcolor{good}Yes & \cellcolor{good}Yes \\
\textit{Space Requirement}         & \cellcolor{good}Small & \cellcolor{good}Small & \cellcolor{mediumgood}Medium & \cellcolor{bad}Large & \cellcolor{mediumbad}Medium Large & \cellcolor{good}Small (0.5 m$^2$ footprint) \\
\textit{Cost / Setup Complexity}   & \cellcolor{good}Very Low & \cellcolor{mediumgood}Low & \cellcolor{mediumgood}Low & \cellcolor{mediumgood}Medium & \cellcolor{bad}Very High & \cellcolor{good}Low (\textless\$250) \\
\bottomrule
\end{tabular}%
}
\end{table*}

Our findings show that \textit{LocoScooter} offers comparable performance to joystick navigation while significantly enhancing immersion, enjoyment, and hedonic quality. Despite higher reported physical demand, participants did not feel more fatigued, suggesting that familiar and sensorimotor-aligned movement can be perceived as part of the interaction rather than as a burden. These findings highlight the potential of compact, metaphor-driven locomotion systems to support expressive, deployable, and sustainable VR interaction. Our key contributions are:

\begin{itemize}
    \item We introduce \textit{LocoScooter}, an affordable, compact locomotion interface. Users steer with a handlebar and adjust speed and direction by sliding their feet. Detailed assembly instructions are also included.

    \item We report findings from a within-subject study ($N = 14$) comparing \textit{LocoScooter} to joystick-based locomotion, revealing enhanced immersion and engagement with comparable usability and performance.

    \item We analyze the dissociation between physical demand and perceived fatigue, illustrating how sensorimotor alignment and meaningful movement may integrate bodily effort into immersive interaction.

    \item We discuss implications for designing embodied locomotion in space-constrained contexts, emphasizing real-world metaphors, directional control, and deployment feasibility.
\end{itemize}

\section{Related Work}

We review prior work across three areas relevant to the design of \emph{LocoScooter}: (1) locomotion techniques in VR, (2) foot-based interaction, and (3) interaction methods tailored to confined physical spaces. This review helps position our system at the intersection of embodied locomotion and compact, deployable interfaces.

\subsection{Locomotion Techniques in VR}

Locomotion is a foundational element of VR interaction, shaping user immersion, presence, and comfort \cite{hale2014handbook}. Traditional joystick-based navigation is simple and widely supported, but it often leads to spatial disorientation and cybersickness due to the lack of congruent bodily movement \cite{lathrop2002perceived,ruddle2006efficient}. To address this, researchers have explored more physically grounded alternatives.

Real walking provides high spatial fidelity but is constrained by the available physical area \cite{wilson2016vr,nabiyouni2015comparing}. Redirected walking techniques subtly alter a user's path to fit larger virtual spaces into limited physical areas \cite{razzaque2005redirected}, though their effectiveness is bounded by perceptual thresholds. Walk-in-place (WIP) techniques offer a space-efficient compromise by simulating forward movement through stepping gestures \cite{bozgeyikli2016locomotion}, yet they may reduce realism and increase fatigue.

Omnidirectional treadmills and systems like the VirtuSphere \cite{calandra2018arm, nabiyouni2015comparing} support natural walking in place but are often costly, bulky, or physically demanding. Other alternatives such as arm-swinging locomotion \cite{mccullough2015myo},  head-based controls like the Human Joystick \cite{mcmahan2012evaluating, kitson2017comparing}, \rev{and leaning-based balance interfaces \cite{zhao_2023_leanon}} reduce hardware requirements but can interfere with simultaneous interaction or provide limited directional precision.

Point-and-teleport techniques offer simplicity and low cybersickness risk \cite{bozgeyikli2019locomotion, cherep2020spatial}, but they break the continuity of movement, reducing presence. Hybrid techniques like World-in-Miniature (WIM) \cite{berger2018wim} help users build mental maps by offering a global perspective, though they require metaphor shifts and interrupt navigation flow.

\begin{figure*}
    \centering
    \includegraphics[width=\linewidth]{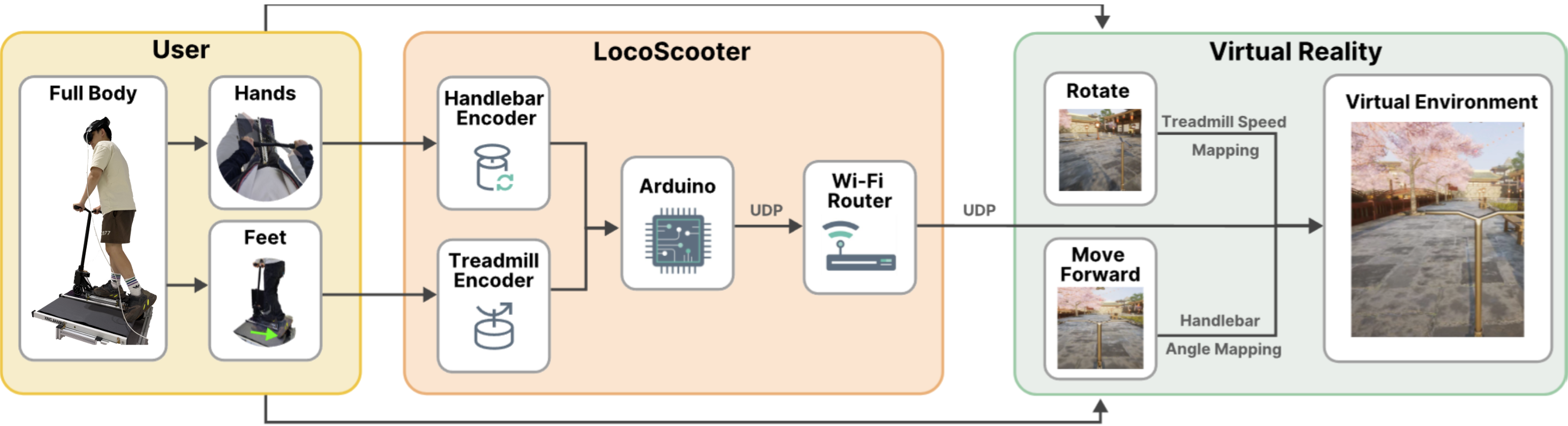}
    \caption{\emph{LocoScooter} system diagram consists of three parts: User, Sensors, and VR. The User section represents the direct interaction through hands and feet, where users rotate the handlebar or step on the treadmill to control the virtual scooter’s forward movement or rotation in the VR environment. The Sensors, equipped with two angle encoders attached to the \emph{LocoScooter}, directly receive the user’s raw action data, which is then analyzed and converted into electric numeric data by a control unit. This processed data is transmitted via UDP through a Wi-Fi router to the VR system. The VR system is responsible for receiving the processed data from the Sensors and mapping it into the avatar’s motion, providing visual cue feedback to the user, which can be observed in the VR.}
    \label{fig:system}
\end{figure*}

\begin{figure}[t]
    \centering
    \includegraphics[width=0.85\linewidth]{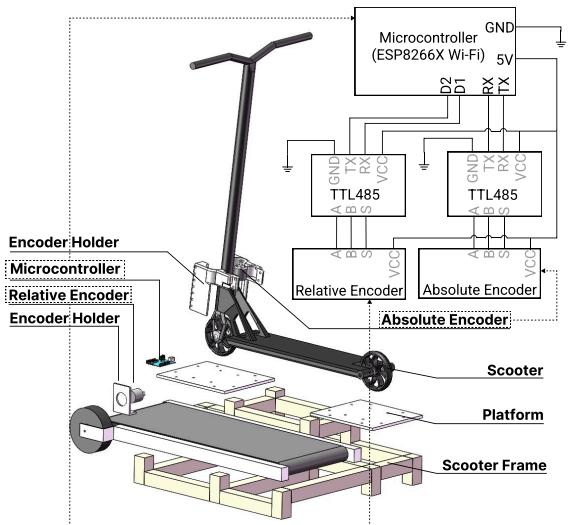}
    \caption{System components of \textit{LocoScooter}. The platform combines a welded frame, commercial scooter, and treadmill. Steering and foot-sliding are tracked by rotary encoders, with signals processed by an ESP32 microcontroller and transmitted wirelessly to the VR application.}
    \label{fig:System_circuit}
\end{figure} 

\subsection{Foot-Based Interaction in VR}

Foot-based interaction leverages the natural role of the lower limbs in human movement and has gained attention for its potential in immersive systems \cite{von2020podoportation, Velloso_2015_feet, lacquaniti2012patterned}. By engaging proprioceptive and biomechanical feedback, these interfaces offer enhanced spatial control and embodied awareness \cite{drossis2013comparative}. Step-based and toe-driven systems have demonstrated notable cognitive and motor advantages; for example, Augsten et al. showed that step-based input reduces mapping errors \cite{augsten2010multitoe}, while Müller et al. found comparable performance between toe-based and hand-based input \cite{muller2023tictactoes}.

Various sensing approaches have been explored, including depth cameras \cite{jota2014let}, pressure-sensitive insoles \cite{scott2010sensing}, and multimodal systems combining foot input with gestures or gaze \cite{sangsuriyachot2011novel}. These systems offer high spatial accuracy and fast learning curves \cite{nilsson2018natural}, often at lower cost. However, many suffer from ergonomic challenges such as fatigue or lack of continuous directional control \cite{trombley1966experimental}. Perceptual redirection methods aim to optimize spatial use \cite{cherep2020spatial}, but often decouple motion from direct physical feedback. 

\subsection{Interaction in Confined Physical Spaces}

With the rise of portable and standalone VR headsets, immersive systems are increasingly used in compact environments such as homes, vehicles, and public spaces \cite{tseng2023fingermapper, williamson2019planevr, hock2017carvr, mcgill2017passenger}. These scenarios impose spatial constraints that limit full-body movement and demand compact, expressive interaction methods~\cite{floyd2021limited}. Prior work has explored interaction in these spaces using spatial redirection, gesture remapping, or miniaturized input. For example, FingerMapper \cite{tseng2023fingermapper} allows full-body interaction via redirected finger motions, and seated VR systems for vehicles use arm-based gestures or passive haptics to maintain usability in motion-constrained settings \cite{mcgill2019virtual, mcgill2020challenges}. While effective for object manipulation and selection tasks, these approaches rarely address navigation or spatial locomotion. This creates a gap for compact locomotion interfaces that maintain embodied control and directional input despite space limitations.

\subsection{Gap and Motivation}

\del{Despite extensive research on VR locomotion, few systems target the specific combination of (1) natural foot-based control, (2) continuous directional navigation, and (3) compatibility with physically constrained environments such as homes, classrooms, or installations. While full-body techniques like real walking and omnidirectional treadmills offer high spatial fidelity, they remain costly, space-demanding, and unsuitable for shared or mobile VR use. Conversely, space-saving techniques like teleportation or joystick control sacrifice embodiment and disrupt motion continuity. Foot-based locomotion interfaces provide a promising middle ground, but existing systems often lack continuous steering or are limited to selection and manipulation tasks rather than sustained navigation.}

\rev{To consolidate the limitations discussed in previous subsections, Table~\ref{tab:locomotion_comparison} compares the key dimensions of existing locomotion techniques. As shown, current systems generally enforce a trade-off: techniques offering high spatial fidelity (e.g., omnidirectional treadmills) are often costly and space-demanding~\cite{razzaque2005redirected,bruno2013new,wilson2016vr,calandra2018arm,warren2017user}, while space-saving methods (e.g., teleportation, joystick) sacrifice embodiment and motion continuity~\cite{chen_2013_navigation,boletsis2019vr}. Although foot-based interfaces provide a promising middle ground~\cite{von2020podoportation, Velloso_2015_feet, lacquaniti2012patterned}, they often lack continuous steering~\cite{trombley1966experimental} or remain limited to specific manipulation tasks. Consequently, there is a clear gap for a system that simultaneously targets (1) natural foot-based control, (2) continuous directional navigation, and (3) compatibility with physically constrained environments~\cite{tseng2023fingermapper, williamson2019planevr, hock2017carvr, mcgill2017passenger}. \textit{LocoScooter} addresses this need by} introducing a compact, scooter-inspired system that blends foot-sliding propulsion with handlebar steering, providing immersive navigation grounded in a real-world movement metaphor. Unlike prior solutions that depend on custom or expensive hardware \cite{cherni2020literature,cannavo2020evaluation}, \textit{LocoScooter} is constructed entirely from off-the-shelf components and designed for fast, low-cost integration into diverse VR setups, including exhibition booths, fitness games, and seated or standing public installations.

\section{LocoScooter: System Design}

\emph{LocoScooter} is a compact, scooter-inspired locomotion interface designed to support embodied navigation in physically constrained VR environments such as public exhibitions, educational setups, or fitness games. By translating familiar physical gestures, foot sliding and handlebar steering into continuous virtual motion, the system enables full-body control over direction and speed while keeping users safely stationary. Rather than replacing general-purpose VR locomotion systems, our goal is to offer a lightweight and deployable option tailored to scenarios where mobility, spatial realism, and physical engagement are desired but room-scale tracking or large installations are impractical.

Several prior systems have explored scooter-like metaphors for VR travel~\cite{lin2024framework,deligiannidis2006vr,sato2015vibroskate}. \textit{VR Scooter}~\cite{deligiannidis2006vr} and \textit{VibroSkate}~\rev{\cite{sato2015vibroskate}} demonstrated the potential of leveraging real-world body motions for immersive navigation. However, these early systems were constrained by the limitations of their sensing and computing technologies. For example, VR Scooter relied on magnetic tracking and mechanical rotation, which introduced latency and resolution issues, while VibroSkate lacked directional steering and required repeated foot tapping for propulsion, reducing overall control and fluidity.

\rev{\emph{LocoScooter} advances the design space of compact, foot-based embodied locomotion, which is defined by the intersection of continuous directional control and stationary physical deployment, with three key contributions.} First, we introduce a dual-modality control scheme that separates directional steering (via handlebar rotation) from forward propulsion (via foot sliding), allowing precise and continuous control within a minimal footprint (see Figure~\ref{fig:system}). Second, we incorporate modern sensing components, high-resolution rotary encoders and low-latency wireless micro-controllers, enabling responsive input and modular integration with contemporary VR engines. Third, we provide a systematic user study evaluating the experiential qualities of scooter-style embodied locomotion in comparison to joystick-based input, targeting task performance, workload, and user satisfaction.

The system is purposefully built to remain affordable (under \$250, see Table~\ref{tab:cost}), compact, and easy to assemble using readily available components. Inspired by the physical stance of real-world scooter riding, the design encourages asymmetric foot placement and coordinated motion to reduce fatigue and promote engagement. In the following sections, we detail the interaction mechanics, hardware architecture, and communication pipeline that make \emph{LocoScooter} a deployable and expressive solution for constrained VR contexts. The complete prototype was developed and assembled within two days with assistance from two professional engineers.

\begin{table}[t]
\caption{Detailed Cost Breakdown for \emph{LocoScooter}}
\label{tab:cost}
\centering
\resizebox{\linewidth}{!}{%
\begin{tabular}{lccc}
\toprule
\textbf{Product Name} & \textbf{Unit Price (USD)} & \textbf{Quantity} & \textbf{Total (USD)} \\
\midrule
 Electronic Component & 9.975 & 4 & 39.9  \\
Router 4A Gigabit Edition & 13.8 & 1 & 13.8 \\
 Metal Universal Joint & 7.49 & 2 & 14.98  \\
 Original USB to Type-C Data Cable & 1.67 & 1 & 1.67  \\
 Indoor Fitness Equipment &  36.12 & 1 & 36.12  \\
 Sports Equipment \& Accessories (Scooter) & 92.12 & 1 & 92.12 \\
 Drawing \& Measuring Equipment & 19.04 & 2 & 38.08  \\ \midrule
 \textbf{Sum} &  & & \textbf{229.7}\\ 
\bottomrule
\end{tabular}%
}
\end{table}

\subsection{Interaction Mechanics}

\begin{figure}[t]
    \centering
    \includegraphics[width=\linewidth]{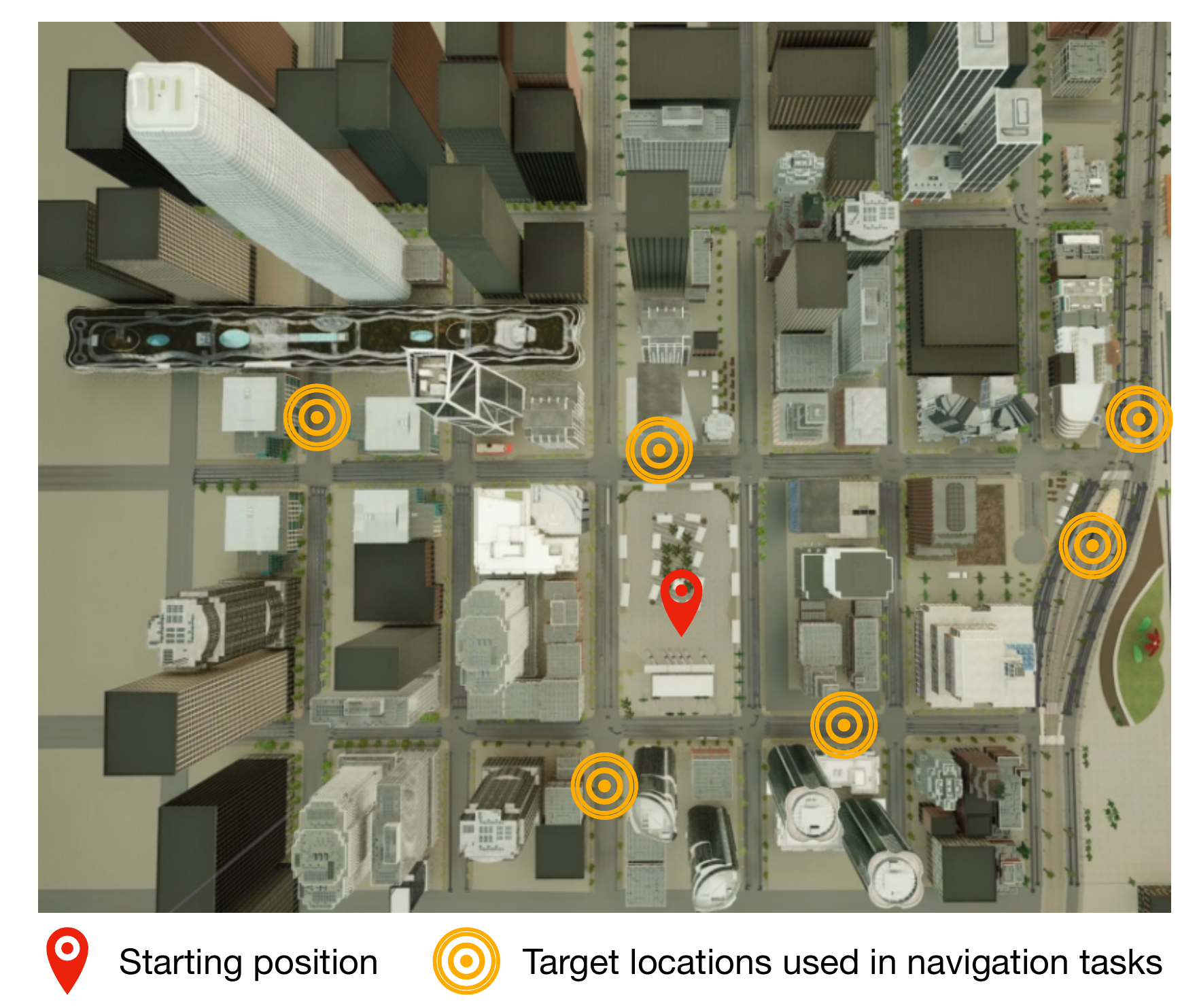}
    \caption{Top-down view of the VR city environment (map) used in the study. The red pin indicates the starting position, and the orange circles represent target locations that participants navigated to during the tasks.}
    \label{fig:map}
\end{figure}

\emph{LocoScooter} enables users to control movement through full-body input, combining foot-sliding with handlebar steering. The user stands with one foot stabilized on the scooter deck and the other on a low-profile treadmill. As the rear foot slides, the system captures both sliding speed and steering angle via a custom-built UDP communication protocol. The raw input is received as a two-dimensional array: \texttt{[angle, speed]}. \rev{In terms of input mapping, \emph{LocoScooter} adopts a decoupled control scheme in which distinct physical inputs are assigned to different locomotion degrees of freedom. Foot-sliding controls forward and backward translational motion, while handlebar rotation controls yaw rotation of the virtual avatar. Consistent with real-world scooter operation, sliding the rear foot forward on the treadmill drives the encoder in one direction and results in forward virtual motion, whereas sliding the foot backward produces backward translation. Yaw rotation is continuously controlled through handlebar rotation directly instead of the speed. Moreover, the system does not support lateral strafing or vertical movement.}

To ensure stable and responsive motion, both input channels are normalized to the range \([-1, 1]\), forming the input domain for virtual locomotion. These values are then mapped to in-game linear and angular velocities using tunable scaling functions and sensitivity parameters. Specifically, forward speed is derived from the sliding input through a configurable acceleration curve, while yaw rotation is determined by the handlebar’s angular displacement. We clamp both linear and angular acceleration within predefined bounds to ensure smooth control. All control parameters are exposed in Unreal Engine for easy calibration. This interaction model enables continuous, expressive movement with minimal setup requirements.

\subsection{Hardware Architecture and Communication Pipeline}

As shown in Figure~\ref{fig:System_circuit}, the \emph{LocoScooter} system consists of a welded steel frame, a commercial scooter with a fixed front wheel, and a compact treadmill mounted behind the deck. The handlebar is attached to a universal joint, allowing rotation for directional input while keeping the base stationary, maintaining stability and enabling an intuitive asymmetric stance.

Two rotary encoders capture user input: an absolute encoder (14-bit resolution) for handlebar rotation and a relative encoder for treadmill motion. These sensors strike a balance between precision and cost, and support long-term durability \rev{(the total hardware cost is approximately \$230; see detailed breakdown in Table~\ref{tab:cost})}. Signals are sampled at 100\,Hz by an ESP32 microcontroller, with RS485 used for absolute position reading and direct GPIO-based counting for relative movement. Data are transmitted over Wi-Fi via UDP to a host PC, allowing low-latency, tether-free integration with the VR system. This wireless communication pipeline supports flexible deployment in shared, mobile, or exhibition contexts \rev{(as detailed in Figure \ref{fig:system})}.

\subsection{Implementation and Compatibility}

The system integrates with Unreal Engine through a custom plugin that maps encoder input into avatar motion. Both direction and speed inputs are normalized and parameterized, enabling adaptation to diverse application contexts, such as walking, skating, or driving metaphors. \emph{LocoScooter} is compatible with HMD-based VR, large displays, and CAVE-style systems. Overall, it balances physical embodiment, directional precision, and ease of deployment, making it a suitable locomotion option for scenarios where compactness, immersion, and physical engagement are desirable.

\section{User Study: Evaluating Embodied Locomotion}

\rev{In this section we elaborate on the evaluation of LocoScooter's experiential qualities.} While joystick control is widely adopted in VR due to its simplicity and familiarity, it often lacks physical engagement and bodily realism. In contrast, \emph{LocoScooter} offers a more sensorimotor-aligned experience by combining foot-sliding and handlebar steering in a compact, metaphor-driven setup. Our study investigates how this embodied input style performs in a directional navigation task designed to reflect real-world scooter movement, focusing on subjective experience and user workload rather than raw efficiency alone.

\paragraph{Joystick Condition.}
\rev{In the joystick condition, we adopted the standard OpenXR locomotion framework\footnote{\href{https://dev.epicgames.com/documentation/en-us/unreal-engine/openxr-input-in-unreal-engine?application_version=5.4}{Epic Games: OpenXR Input in Unreal Engine}}, which is commonly used in commercial VR applications. Navigation was controlled using the handheld controller’s analog joysticks. Forward and backward translation were mapped to the Y-axis of the left-hand joystick, while turning was mapped to the X-axis of the right-hand joystick. For directional control, only the forward-facing half of the joystick’s range (i.e., the upper $180^\circ$) was used. Left–right joystick deflection was linearly mapped to virtual yaw rotation, corresponding directly to the handlebar rotation in the \textit{LocoScooter} condition, with a one-to-one angular mapping from $-90^\circ$ to $+90^\circ$. OpenXR normalizes joystick input to the range $[-1, 1]$ by default. These normalized values were passed through the same control pipeline used for \textit{LocoScooter}, including linear sensitivity scaling and clamped acceleration. To ensure a fair comparison, both conditions shared identical acceleration parameters and maximum velocity limits at the engine level, resulting in matched virtual motion profiles. This design isolates the effect of physical embodiment from differences in underlying virtual locomotion dynamics.
}
\begin{figure*}
    \centering
    \includegraphics[width=\linewidth]{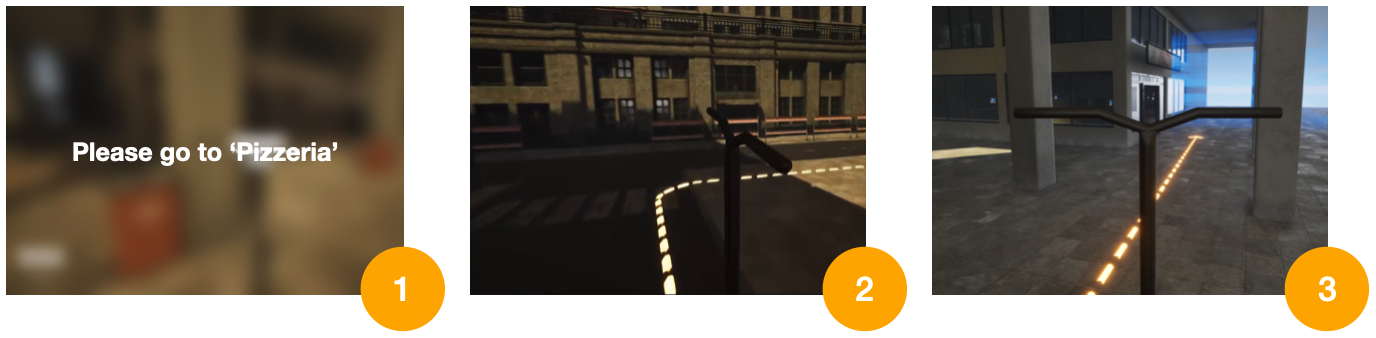}
    \caption{Illustration of the experimental procedure. (1) Participants receive a location prompt (e.g., ``Please go to 'Pizzeria'"). (2) A navigation path appears on the ground in the virtual environment to guide movement. (3) Upon reaching the blue goal zone, participants must stop and dwell for 2 seconds to complete the round. They are then teleported back to the starting position to begin the next of six trials, each targeting a different location.}
    \label{fig:procedure}
\end{figure*}

\begin{figure}[t]
    \centering
    \includegraphics[width=\linewidth]{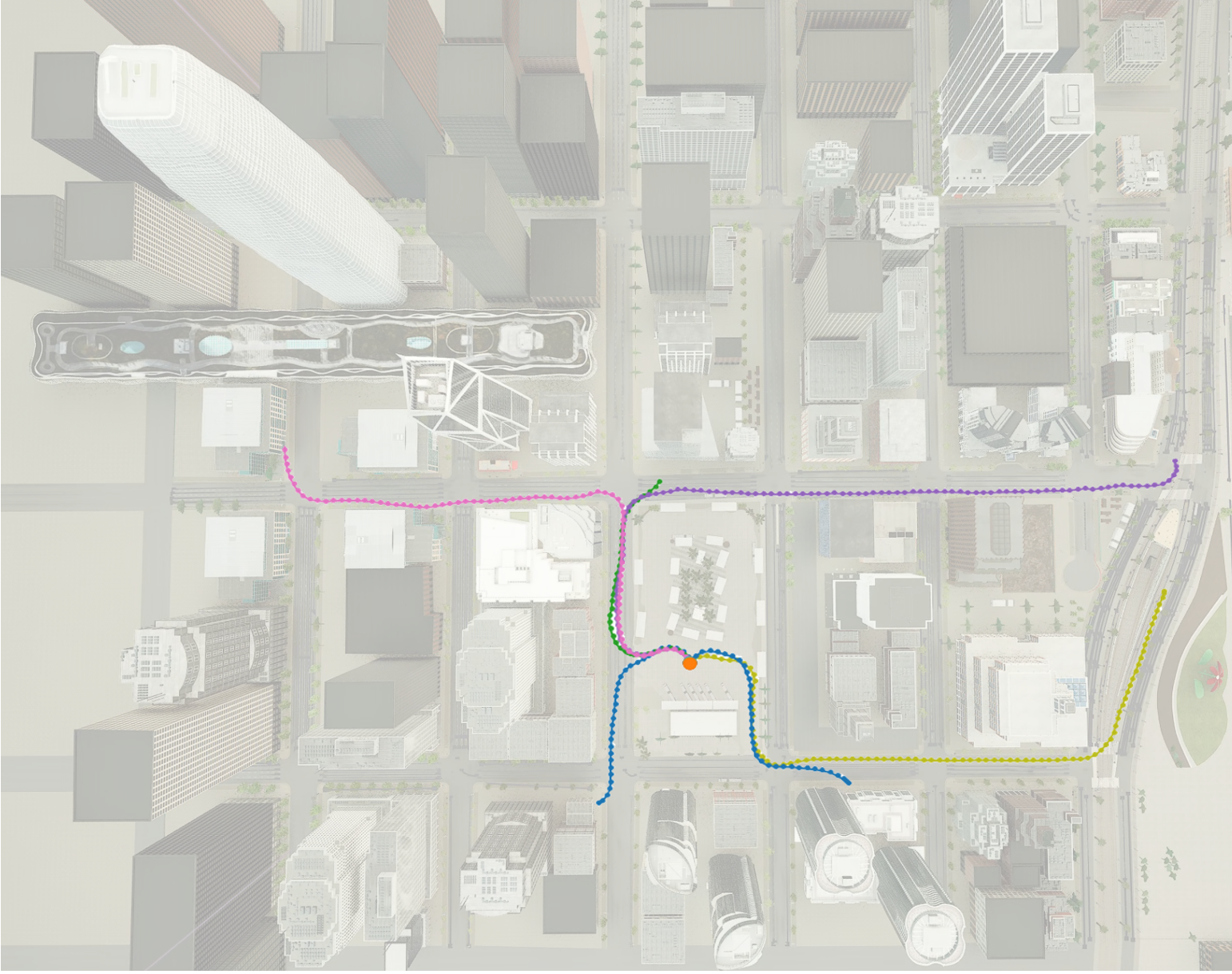}
    \caption{Representative paths taken by participants in the virtual city navigation task. Colored lines illustrate different routes from the starting location (orange dot) to various target destinations across the environment. This visualization highlights the spatial coverage and directional variety of participant movements during the experiment.}
    \label{fig:guide-map}
\end{figure}

\begin{figure}[t]
    \centering
    \includegraphics[width=\linewidth]{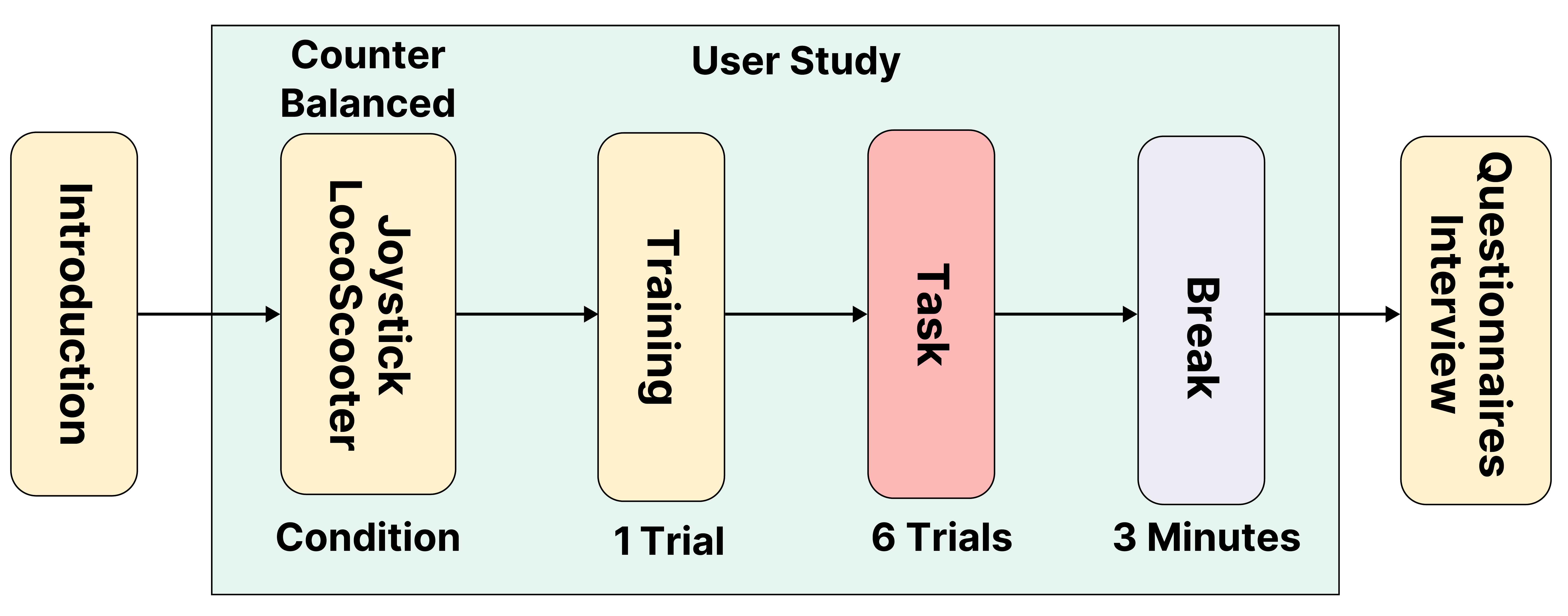}
    \caption{Overview of the user study procedure. Participants experienced both the Joystick and LocoScooter locomotion conditions in a counterbalanced order. Each condition included a training trial, six task trials, and a 3-minute break before completing questionnaires and a semi-structured interview.}
    \label{fig:user_study}
\end{figure}
 
\subsection{Motivation and Research Questions}

Embodied locomotion has been shown to enhance immersion, spatial awareness, and user enjoyment in virtual environments by aligning physical effort with virtual motion \cite{Nybakke_2012_mobility}. While walking-based techniques offer high sensorimotor fidelity, their implementation is often impractical due to spatial or safety constraints \cite{Cirio_2012_walking}. This creates opportunities for embodied alternatives that retain physical involvement while remaining lightweight and deployable.

\textit{LocoScooter} addresses this need by embedding directional control and propulsion within a compact footprint, inspired by real-world scooter riding. It is particularly suited for contexts where users can stand in place and engage with VR content through embodied gestures, such as public demos, interactive installations, or VR fitness modules. However, it remains unclear how such a design affects performance, perceived effort, and overall user satisfaction when compared to standard joystick navigation. \rev{Further, a key open question is whether the familiarity and naturalness of the movement can alter how users perceive their own physical effort---specifically, whether high sensorimotor alignment can dissociate subjective fatigue from objective physical demand.} Therefore, our study aims to answer:

\begin{itemize}
  \item[\textbf{RQ1:}] How does embodied locomotion via \textit{LocoScooter} affect task efficiency and perceived workload in spatial navigation?
  \item[\textbf{RQ2:}] How does the embodied nature of \textit{LocoScooter} shape subjective experience, including immersion, enjoyment, and user satisfaction?
  \item[\textbf{RQ3:}] To what extent can familiar, sensorimotor-aligned movement mitigate the subjective cost of physical exertion?
\end{itemize}

\subsection{Formative Study and Calibration}

To design a task suitable for evaluating directional navigation through a scooter-based interface, we conducted a formative study with seven participants (6 male, 1 female; age range 22-–32; $M = 26.1$, $SD = 3.3$). Participants used both the \emph{LocoScooter} and joystick input to navigate through a large-scale virtual city built in Unreal Engine 5.4.4. The goal was to identify target locations that provided spatial variety while maintaining a manageable task duration to minimize simulator sickness and ensure ecological validity.

We prepared a pool of candidate destinations differing in urban layout, route complexity, and travel distance. Participants traveled to multiple destinations using both input methods. For each route, we recorded traversal time and collected qualitative feedback on effort, disorientation, and perceived comfort. Based on these insights, we selected six final target locations (Figure~\ref{fig:map}) that consistently yielded travel durations between 1–2 minutes, long enough to reflect meaningful navigation tasks, yet short enough to avoid fatigue or motion discomfort. These locations were distributed across different spatial configurations to enable fair within-subject comparison across both input conditions.

\subsection{Participants and Apparatus}

We recruited 14 participants (9 male, 5 female) from a local university, aged 18–-32 years ($M = 25.00$, $SD = 3.06$). All participants had normal or corrected-to-normal vision and no known motor impairments. None of them had participated in the formative study. Prior VR experience was self-reported on a 7-point Likert scale ($M = 4.79$, $SD = 1.63$), ranging from minimal to advanced experience. Motion sickness susceptibility was assessed using the MSSQ-Short~\cite{golding1998motion}; no participants scored in the high-risk range. Each session lasted approximately 45 minutes, and participants received a small gift card as compensation. \rev{The study protocol was approved by the Institutional Review Board under approval ID HKUST(GZ)-HSP-2025-0097.}

The experiment was run on a PC equipped with an RTX-4090 GPU and Unreal Engine 5.4.4, with visuals streamed to a Meta Quest Pro via Oculus Link. The \emph{LocoScooter} interface transmitted input wirelessly, while the joystick condition used the right-hand controller. All virtual environments, routes, and visual cues were identical across conditions. Participants performed the locomotion tasks while standing and completed questionnaires while seated. This setup reflects the envisioned deployment scenarios of \emph{LocoScooter}—short-burst, physically engaging interactions in compact spaces such as exhibitions, demonstrations, or fitness stations.

\subsection{Study Design and Procedure}

We conducted a within-subject experiment to compare \textit{LocoScooter} with conventional joystick-based locomotion in a goal-oriented navigation task. The study followed a counter-balanced design to mitigate ordering effects. Each participant completed the same set of tasks using both locomotion methods, with condition order randomized across participants.

\begin{figure*}
    \centering
    \includegraphics[width=0.9\linewidth]{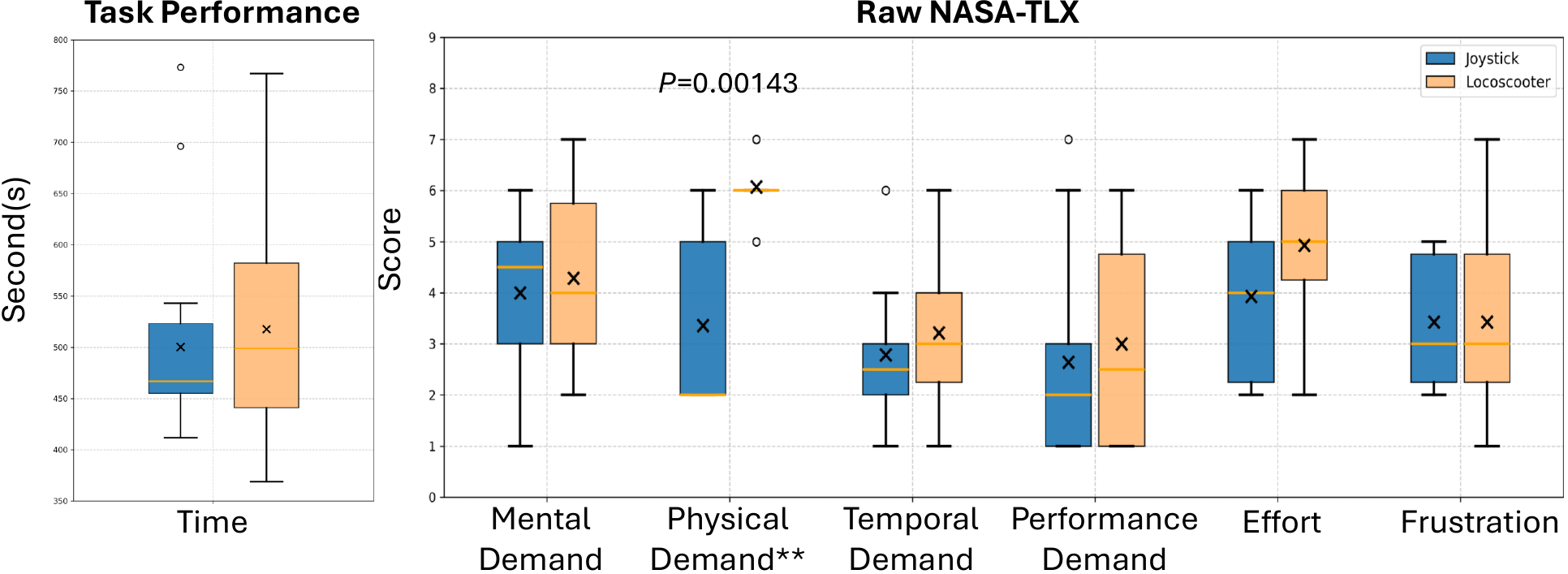}
    \caption{Overview of task performance and perceived workload across the two locomotion conditions. Left: Task completion time (in seconds); Right: Raw NASA-TLX subscale scores. Orange lines indicate the median, black crosses (\(\times\)) indicate the mean, and dots (\(\circ\)) represent individual outliers. Asterisks indicate significance levels: * \(p < .05\), ** \(p < .01\), *** \(p < .001\). }
    \label{fig:nasa-task}
\end{figure*}

Participants were immersed in a large-scale virtual city and instructed to navigate to a sequence of target locations using either the \emph{LocoScooter} or joystick. Each trial began with an on-screen textual prompt (e.g., “Please go to ‘Pizzeria’”), after which a visual navigation path appeared in the scene to guide the participant (see Figure~\ref{fig:procedure}(1)). Participants followed the path using their assigned input method and stopped within a blue-highlighted target zone to complete the trial. To ensure consistency in endpoint detection, participants were required to remain stationary within the zone for two seconds before the system progressed to the next round (see Figure~\ref{fig:procedure}(2)–(3)). Between trials, participants were automatically returned to a common starting point via teleportation.

Before beginning the formal study, each participant completed a brief training session (about 1 minute) with both input methods to familiarize themselves with the system. \rev{Specifically for \emph{LocoScooter}, to address the lack of hand tracking, we ensured precise visuo-haptic correspondence between the physical handlebar and its virtual counterpart. During training, participants were instructed to familiarize themselves with this alignment, allowing them to reliably locate and grasp the handlebar using proprioception.} During the main study, each locomotion condition included six randomized navigation trials drawn from the calibrated pool of destinations established in the formative study (Figure~\ref{fig:guide-map}). These targets were selected to provide diversity in spatial distribution, route length, and directional complexity, while maintaining navigational realism. The randomized trial order further reduced memorization and allowed us to assess each condition across a range of urban navigation subtasks. \rev{Upon completion of all conditions, we conducted a semi-structured post-study interview to gather qualitative feedback regarding participants' user experience and design suggestions.}

In total, each participant completed 12 navigation trials: 2 conditions $\times$ 6 destinations. Across all 14 participants, this yielded 168 trials overall (see Figure \ref{fig:user_study}). The user's viewpoint in VR was fixed to a virtual scooter model throughout the study, providing consistent visual feedback and contextual anchoring across both conditions.

\subsection{Measures}

We collected both objective and subjective measures to evaluate locomotion effectiveness and user experience. 
Task performance was measured by total completion time (in seconds). 
Perceived workload was assessed using the Raw NASA-TLX~\cite{hart1988development}. \rev{We specifically distinguish Physical Demand, which measures the instantaneous intensity required to execute the task (e.g., rotating, sliding), from Physical Fatigue.} Perceived fatigue level was measured with the Borg RPE scale (6-–20)~\cite{borg1982psychophysical} \rev{to evaluate the user's cumulative physiological exhaustion state after completing the task}. 
To assess immersion, we used the Igroup Presence Questionnaire (IPQ)~\cite{schubert2001experience}, while overall user experience was measured using the UEQ-Short~\cite{schrepp2017ueq}. 
We also evaluated the system usability via the System Usability Scale (SUS)~\cite{brooke1996sus}, and enjoyment was rated on a single 7-point Likert item. Finally, simulator sickness was assessed using the Simulator Sickness Questionnaire (SSQ)~\cite{kennedy1993simulator}. \rev{All subjective measures were administered immediately following the completion of each experimental condition to capture the participants' immediate experiential feedback.}

\subsection{Results}

\begin{figure*}[h]
    \centering
    \includegraphics[width=0.9\linewidth]{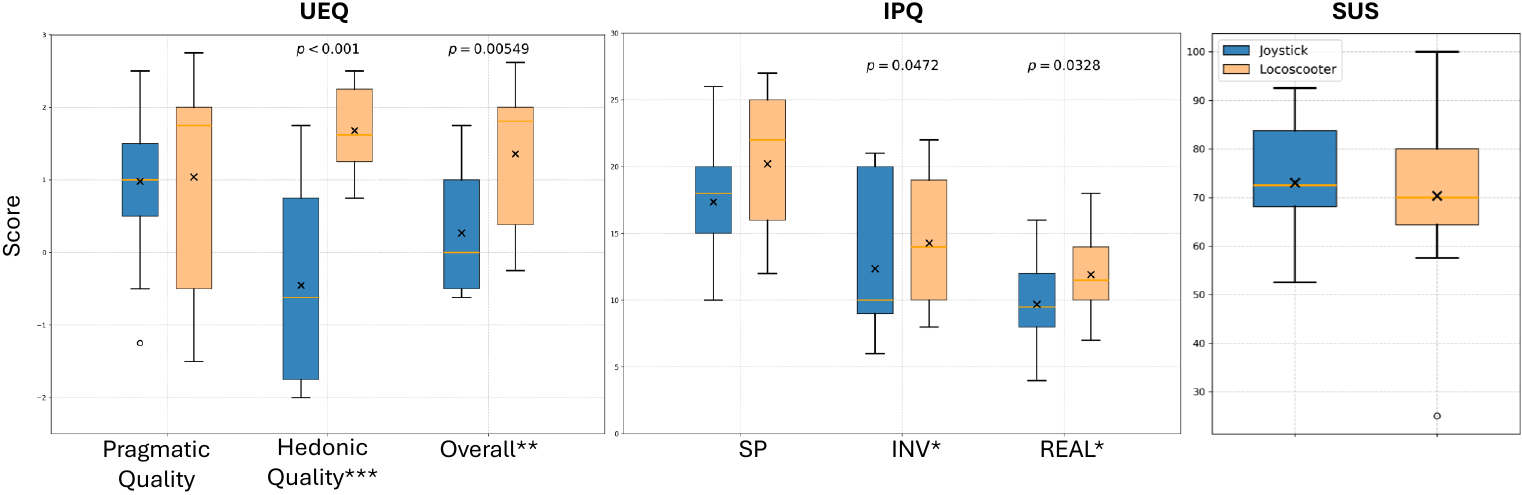}
    \caption{Overview of subjective experience ratings across the two locomotion conditions. Left: User Experience Questionnaire (UEQ) scores for pragmatic quality, hedonic quality, and overall impression. Middle: Immersion measures from the Igroup Presence Questionnaire (IPQ), including spatial presence (SP), involvement (INV), and realism (REAL). Right: System Usability Scale (SUS) scores. Orange lines indicate the median, black crosses (\(\times\)) indicate the mean, and dots (\(\circ\)) represent individual outliers. Asterisks indicate significance levels: * \(p < .05\), ** \(p < .01\), *** \(p < .001\).}
    \label{fig:uew-ipq-sus}
\end{figure*}

\begin{figure*}[h]
    \centering
    \includegraphics[width=0.9\linewidth]{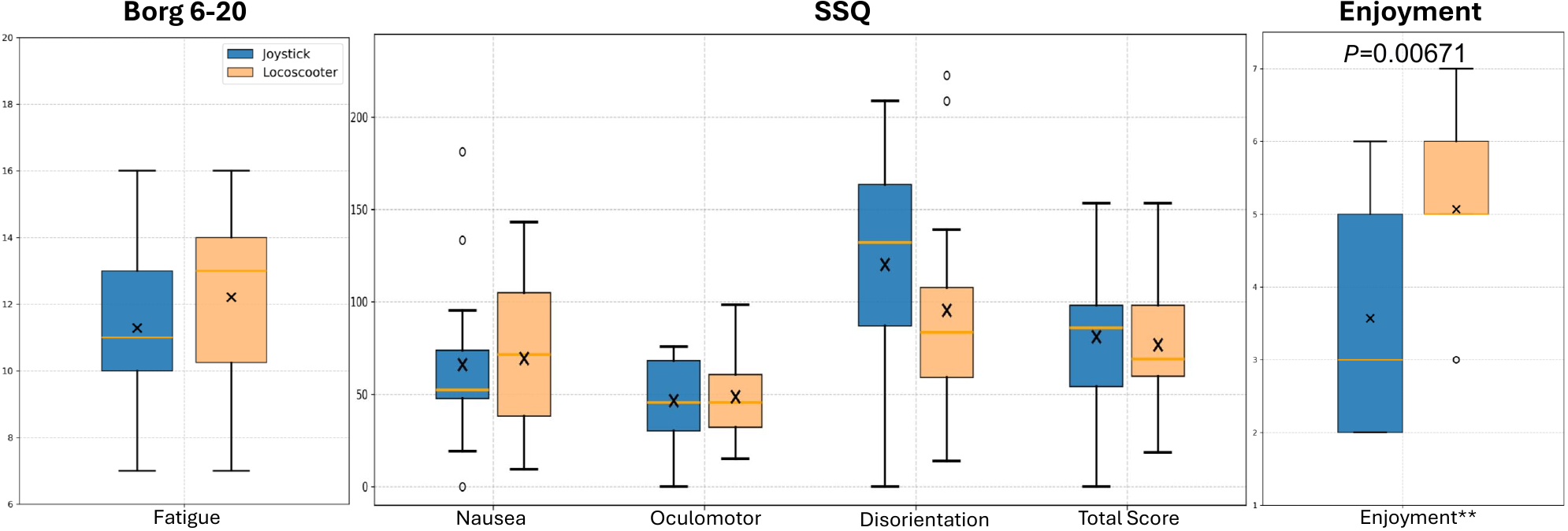}
    \caption{Overview of perceived fatigue, simulator sickness, and enjoyment across the two locomotion conditions. Left: Borg RPE (6–20 scale) scores for perceived physical fatigue. Middle: Simulator Sickness Questionnaire (SSQ) scores, including subscales for nausea, oculomotor symptoms, and disorientation, as well as total score. Right: Enjoyment ratings. Orange lines indicate the median, black crosses (\(\times\)) indicate the mean, and dots (\(\circ\)) represent individual outliers. Asterisks indicate significance levels: * \(p < .05\), ** \(p < .01\), *** \(p < .001\).}
    \label{fig:enjoyment-fatigure}
\end{figure*}

We analyzed the effects of locomotion technique (i.e., \textit{LocoScooter} vs. \textit{Joystick}) on all dependent measures using a within-subject design. To compare the two experimental conditions, we conducted separate analyses for each dependent variable. We first assessed the normality of the distribution of the differences between paired observations using the Shapiro--Wilk test. If the normality assumption was met, we performed a paired-samples $t$-test. If the assumption was violated, we used the non-parametric Wilcoxon signed-rank test. All statistical tests were evaluated using an alpha level of .05. For all significant results, we report the corresponding effect sizes (i.e., Cohen’s $d$ for $t$-tests, or rank-biserial $r$ for Wilcoxon tests). The results are summarized in Figures~\ref{fig:nasa-task}, \ref{fig:uew-ipq-sus}, and \ref{fig:enjoyment-fatigure}.

\subsubsection{Task Performance (Time)}

\paragraph{Task Completion Time.}
\textit{LocoScooter} ($M = 517.7$\,s, $SD = 129.1$\,s) did not differ significantly from \textit{Joystick} ($M = 500.4$\,s, $SD = 116.7$\,s) in task completion time ($t(13) = 1.08$, $p = .300$).

\subsubsection{Perceived Workload (NASA-TLX)}

\paragraph{Physical Demand.}
\textit{LocoScooter} ($M = 6.07$, $SD = 0.62$) was rated significantly higher than \textit{Joystick} ($M = 3.36$, $SD = 1.69$) in physical demand ($W = 91$, $p = .001$, $r = 0.64$).

\paragraph{Other TLX Dimensions.}
No significant differences were found between \textit{LocoScooter} and \textit{Joystick} in mental demand ($t(13) = 0.54$, $p = .598$), temporal demand ($W = 28.5$, $p = .143$), performance ($W = 13$, $p = .672$), effort ($t(13) = 2.13$, $p = .053$), or frustration ($W = 57$, $p = .797$).

\subsubsection{User Experience (UEQ)}

\paragraph{Hedonic Quality.}
\textit{LocoScooter} ($M = 1.68$, $SD = 0.55$) was rated significantly higher than \textit{Joystick} ($M = -0.45$, $SD = 1.24$) in hedonic quality ($t(13) = 6.15$, $p < .001$, $d = 1.64$).

\paragraph{Overall UEQ Score.}
\textit{LocoScooter} ($M = 1.36$, $SD = 0.92$) was rated significantly higher than \textit{Joystick} ($M = 0.27$, $SD = 0.84$) in overall user experience ($t(13) = 3.32$, $p = .005$, $d = 0.89$).

\paragraph{Pragmatic Quality.}
No significant differences were found in pragmatic quality between conditions ($t(13) = 0.12$, $p = .910$).

\subsubsection{Immersion (IPQ)}

\paragraph{Spatial Presence (SP).}
\textit{LocoScooter} ($M = 20.21$, $SD = 5.49$) did not differ significantly from \textit{Joystick} ($M = 17.36$, $SD = 4.41$) in spatial presence ($t(13) = 2.03$, $p = .064$).

\paragraph{Involvement (INV).}
\textit{LocoScooter} ($M = 14.29$, $SD = 4.78$) was rated significantly higher than \textit{Joystick} ($M = 12.36$, $SD = 5.60$) in involvement ($t(13) = 2.19$, $p = .047$, $d = 0.59$).

\paragraph{Realism (REAL).}
\textit{LocoScooter} ($M = 11.93$, $SD = 3.15$) was rated significantly higher than \textit{Joystick} ($M = 9.71$, $SD = 3.15$) in realism ($t(13) = 2.39$, $p = .033$, $d = 0.64$).

\subsubsection{System Usability (SUS)}

\textit{Joystick} ($M = 73.04$, $SD = 14.01$) and \textit{LocoScooter} ($M = 70.36$, $SD = 17.45$) did not differ significantly in system usability scores ($t(13) = -0.39$, $p = .700$). Both scores fall within the range typically considered ``good'' usability (i.e., above 68), suggesting that participants perceived both locomotion methods as acceptably usable~\cite{brooke1996sus}.

\subsubsection{Perceived Fatigue}

\paragraph{Physical Fatigue (Borg RPE 6-20).}
No significant difference in physical fatigue was observed between \textit{LocoScooter} and \textit{Joystick} ($t(13) = 1.15$, $p = .269$).

\subsubsection{Simulator Sickness (SSQ)}

\begin{table}[h]
\centering
\caption{Descriptive statistics for SSQ subscales by condition.}
\label{tab:ssq-subscales}
\begin{tabular}{lcc}
\toprule
\textbf{Subscale} & \textbf{Joystick} ($M \pm SD$) & \textbf{LocoScooter} ($M \pm SD$) \\
\midrule
Nausea        & $66.10 \pm 46.05$ & $69.51 \pm 40.38$ \\
Oculomotor    & $46.56 \pm 23.76$ & $48.73 \pm 24.47$ \\
Disorientation& $120.31 \pm 61.92$ & $95.45 \pm 60.27$ \\
\bottomrule
\end{tabular}%
\end{table}

\paragraph{Total SSQ Score.}
\textit{LocoScooter} ($M = 76.94$, $SD = 40.09$) did not differ significantly from \textit{Joystick} ($M = 81.21$, $SD = 41.19$) in total simulator sickness ($t(13) = -0.34$, $p = .738$). Descriptive statistics for the three SSQ subscales (i.e., nausea, oculomotor, and disorientation) are summarized in Table~\ref{tab:ssq-subscales}, showing comparable symptom patterns across conditions.

\subsubsection{Enjoyment}

\paragraph{Enjoyment.}
\textit{LocoScooter} ($M = 5.07$, $SD = 1.33$) was rated significantly higher than \textit{Joystick} ($M = 3.57$, $SD = 1.40$) in enjoyment ($t(13) = 3.22$, $p = .007$, $d = 0.86$).

\subsubsection{Semi-structured Interview}

We conducted a thematic analysis of post-study interviews and identified four recurring themes: (1) \textit{Perceived Effort and Fatigue}, (2) \textit{Immersion and Sensorimotor Coupling}, (3) \textit{Usability and Learnability}, and (4) \textit{Design Feedback and Future Applications}.

\paragraph{Perceived Effort vs. Fatigue.}
Participants widely acknowledged that \textit{LocoScooter} required greater bodily involvement than joystick control, particularly due to continuous foot-sliding and upper-body coordination. However, this increased physicality did not lead to elevated subjective fatigue. Several participants described the effort as ``acceptable and even enjoyable'' (P15), or ``natural enough that I didn’t feel tired'' (P9). \del{This reinforces our quantitative observation that embodied movement — when sensorimotor-aligned — can support engaging interaction without compromising user comfort.}

\paragraph{Immersion and Sensorimotor Coupling.}
Many participants reported a heightened sense of immersion with \textit{LocoScooter}, attributed to the close alignment between physical movement and virtual motion. P14 remarked that it felt ``more real, like actually moving through space,'' while P6 noted that ``the effort matched the movement, which made it satisfying.'' Several also felt that this congruence reduced disorientation and motion discomfort, especially during forward navigation. A few participants (P6, P12) mentioned that sharp turns or downhill slopes sometimes caused perceptual mismatch, suggesting opportunities for refining visual feedback or tilt simulation.

\paragraph{Usability and Learnability.}
Joystick input was consistently seen as easier to learn and more forgiving, especially for beginners. However, \textit{LocoScooter} was described as intuitive by users familiar with real-world scooters. P11 stated that ``foot-and-handlebar control felt more logical than pressing buttons,'' and P8 mentioned that it ``started to feel natural after just a few minutes.'' Some participants experienced minor difficulty with turning precision or lateral balance, indicating areas for mechanical or software tuning.

\paragraph{Design Feedback and Future Applications.}
Participants offered concrete suggestions for improvement, including adjustable handlebar height (P2), enhanced responsiveness during steering (P10), and additional feedback modalities such as vibration or tilt (P6, P12). Others proposed new application scenarios, such as exercise-based VR or guided exploration. P4 suggested, ``I would totally use this to navigate a virtual museum, as long as the motion feels smooth.'' These responses highlight both the current viability and future potential of scooter-style embodied locomotion in VR experiences that prioritize presence, movement, and compact setup.

\section{Discussion}

Our study examined \textit{LocoScooter}, a compact embodied locomotion interface combining foot-sliding and handlebar steering, in comparison to conventional joystick-based input. We assessed its effects across task performance, user experience, and physical ergonomics, addressing \textbf{RQ1--RQ3}.

In response to \textbf{RQ1}, we found no significant difference in task completion time between \textit{LocoScooter} and joystick input, suggesting that embodied locomotion can offer efficiency comparable to symbolic control. Although \textit{LocoScooter} demanded greater physical effort, it did not increase perceived mental workload or frustration. For \textbf{RQ2}, participants reported significantly higher immersion, enjoyment, and hedonic quality with \textit{LocoScooter}, highlighting the experiential benefits of full-body, metaphor-driven input. Despite its embodied nature, usability scores remained on par with joystick control. Addressing \textbf{RQ3}, we identified a dissociation between physical demand and subjective fatigue: although users exerted more effort, they did not report greater exhaustion. This points to the role of meaningful, sensorimotor-aligned movement in mitigating the perceived cost of physical interaction. We elaborate on this below.

Together, these findings suggest that \textit{LocoScooter} can serve as a compelling, deployable alternative to conventional VR locomotion—supporting engaging, physically expressive navigation without compromising usability or comfort.

\subsection{Dissociation Between Physical Demand and Perceived Fatigue}

A key insight from our study is the dissociation between \rev{perceived} physical demand and subjective fatigue. While NASA-TLX scores indicated significantly higher effort for \textit{LocoScooter}, users did not feel more fatigued. \rev{This suggests that participants successfully distinguished between the process of exertion (active engagement) and the outcome of exhaustion (fatigue).} As P15 noted, “It felt more active, but not really exhausting,” and P9 emphasized that the movement felt “natural.” This supports prior findings that embodied actions, when sensorimotor-aligned and metaphorically meaningful, can be cognitively reframed as part of the interaction experience, rather than a source of strain~\cite{kilteni2012sense, dourish2001action,he2025exploration}.

In our case, foot-sliding and handlebar steering closely mirrored a real-world behavior (riding a scooter), enabling intuitive coordination and rhythmic motion. These qualities likely helped users internalize the physical exertion as functional input rather than extraneous load. Moreover, the predictability and visual feedback of movement may have supported tighter sensorimotor coupling, thereby reducing perceived effort~\cite{furuya2018individual}. \rev{Theoretically, this extends the model of \textit{embodied agency}~\cite{kilteni2012sense} by demonstrating that high sensorimotor congruence allows users to reinterpret physical fatigue not as a system limitation, but as a necessary component of realism, effectively shifting the locus of exertion from external workload to internal engagement.}

\subsection{Design Implications}

The findings offer several implications for designing embodied locomotion systems that are compact, expressive, and sustainable in practice. Drawing on empirical results and participant insights, we reflect on how system-level design choices, including metaphor, control mapping, and physical adjustability, can foster bodily engagement without excessive fatigue.

\paragraph{Embed physical input within familiar metaphors.}
Ten of fourteen participants emphasized that the scooter metaphor made interaction intuitive and reduced learning effort. This aligns with higher hedonic quality and enjoyment scores for \textit{LocoScooter}. P11 described the control as “more logical than pressing buttons,” highlighting the benefit of grounding full-body input in familiar schemas. Metaphor-driven design may thus facilitate effort absorption and user onboarding~\cite{Li_2024_metaphors}.

\paragraph{Align sensorimotor input with virtual response.}
Despite increased physical effort, fatigue scores remained stable across conditions (Borg RPE: $p = .269$). Seven participants linked this to strong sensorimotor congruence, e.g., P6 said, “the effort matched the movement.” These observations reinforce the idea that when action and feedback are tightly coupled, physical exertion becomes part of a fluid interaction loop rather than a cognitive burden~\cite{kilteni2012sense}.

\paragraph{Balance engagement and effort to support sustained use.}
Participants expressed willingness to use \textit{LocoScooter} for longer sessions, especially in contexts like education, fitness, or virtual exploration. Enjoyment scores were significantly higher ($p = .007$), and \rev{no increase in reported fatigue was observed over the course of the experiment. Together, these results suggest that the embodied input was well-aligned with the task demands and supported a positive short-term user experience.} For example, P4 envisioned using the system to “navigate a museum in VR,” pointing to applications where presence and motion enhance experience.

\paragraph{Support adjustability for user diversity.}
Six participants requested physical calibration features (e.g., handlebar height, turning stiffness). While these did not significantly affect usability ratings (SUS: $p = .700$), they were frequently cited in interviews as barriers to control fluency. P2 remarked, “I wish I could raise the handlebar a bit.” This underscores the importance of modular design to accommodate individual anthropometry—especially in shared or public installations.

\subsection{Limitations and Future Work}

Despite these promising results, our study has several limitations. The relatively small sample size ($N = 14$) and short-term exposure limit the generalizability of our findings. \rev{Additionally, we did not explicitly survey participants' prior experience with riding physical scooters. Consequently, the potential influence of pre-existing scooter proficiency on the learning curve and usability ratings remains unquantified.} Future work could examine how perceived fatigue and bodily effort evolve with extended use over time, particularly in task contexts requiring prolonged navigation. Moreover, while our findings point to a dissociation between physical demand and perceived fatigue, we relied solely on subjective measures. \rev{Further, our evaluation lacked fine-grained objective performance metrics; in the future, incorporating automated logging of parameters such as turning radius, path drift, and postural stability would provide more informative insights into the precision and control dynamics of scooter-based locomotion.} Incorporating physiological data such as heart rate, muscle activation (EMG), or galvanic skin response could yield deeper insights into the embodied cost and cognitive integration of physical effort.

Participants also identified practical challenges that future work could address. Some users experienced discomfort during turning or balance instability (P6, P12), while others noted that the turning radius felt too wide or difficult to control precisely (P10, P14). \rev{Further, since LocoScooter requires users to keep both hands on the handlebars for steering and stability, it inherently limits the capacity for concurrent manual interactions, such as ray-casting or grasping virtual objects. While the handlebar serves as a physical anchor that mitigates fall risks during blind locomotion, addressing potential safety concerns, this constraint implies that our system is best suited for navigation-heavy scenarios rather than tasks requiring frequent bimanual manipulation. Additionally, our current results are anchored to navigation tasks within a single large virtual city environment; future evaluations across more varied virtual contexts and task types could further clarify the system's specific strengths and constraints.} Several participants (P2, P8) suggested the need for more adjustable or customizable system components, such as handlebar height or foot support calibration. Additionally, users commented on the lack of braking or momentum feedback (P13, P14), which may reduce realism or break immersion in downhill scenarios.

Finally, future studies could explore alternative embodied interaction metaphors beyond scooter-based movement and systematically vary the degree of sensorimotor alignment to better understand how different mappings affect immersion, fatigue, and control across user populations.

\section{Conclusion}

In this paper, we presented \textit{LocoScooter}, a compact, scooter-inspired locomotion interface that enables embodied navigation in VR through foot-sliding and handlebar steering. Designed for confined spaces and built entirely from commodity components, the system provides directionally expressive control without requiring room-scale tracking or specialized hardware.

We evaluated \textit{LocoScooter} in a within-subject study ($N = 14$), using joystick-based navigation as a commercial baseline. Our findings addressed three core questions related to embodied locomotion. First, task efficiency was maintained despite increased physical engagement. Second, users reported significantly greater immersion, enjoyment, and hedonic experience with \textit{LocoScooter}. Third, although physical demand increased, perceived fatigue remained unchanged. This dissociation between effort and subjective cost suggests that embodied input, when well-aligned with user expectations and system feedback, can enhance experience without adding discomfort. These results highlight the potential of sensorimotor-aligned, metaphor-driven interaction to integrate bodily effort into the experience itself. Overall, \textit{LocoScooter} demonstrates how low-cost, physically expressive locomotion interfaces can support engaging, sustainable, and deployable VR interaction across diverse use contexts.

\section*{Open Science}

To support open science, we provide the assembly design for our LocoScooter in the supplemental material.

\acknowledgments{This research is funded by AI Research and Learning Base of Urban Culture under Project 2023WZJD008.
 Xiang Li is supported by the China Scholarship Council (CSC) International Cambridge Scholarship (No. 202208320092).}

\bibliographystyle{abbrv-doi-hyperref}

\bibliography{scooter}

\end{document}